\documentclass[11pt]{article}

\usepackage[totalwidth=460truept,totalheight=600truept]{geometry}
\usepackage{epsfig,latexsym,amssymb}
\usepackage[T1]{fontenc}

\def\theequation{\arabic{section}.\arabic{equation}}
\def\text#1{\hbox{#1}}

\renewcommand{\theequation}{\thesection.\arabic{equation}}
\global\arraycolsep=1truept

\begin{document}

\hfill \hfill IFUP-TH 2007/34

\vskip 1.4truecm

\begin{center}
{\huge \textbf{Weighted Scale Invariant}}

{\huge {\large \vskip .1truecm}}

{\huge \textbf{Quantum Field Theories}}

\vskip 1.5truecm

\textsl{Damiano Anselmi}

\textit{Dipartimento di Fisica ``Enrico Fermi'', Universit\`{a} di Pisa, }

\textit{Largo Pontecorvo 3, I-56127 Pisa, Italy, }

\textit{and INFN, Sezione di Pisa, Pisa, Italy}

damiano.anselmi@df.unipi.it
\end{center}

\vskip 2truecm

\begin{center}
\textbf{Abstract}
\end{center}

\bigskip

{\small We study a class of Lorentz violating quantum field theories that
contain higher space derivatives, but no higher time derivatives, and become
renormalizable in the large }$N${\small \ expansion. The fixed points of
their renormalization-group flows provide examples of exactly ``weighted
scale invariant'' theories, which are noticeable Lorentz violating
generalizations of conformal field theories. We classify the scalar and
fermion models that are causal, stable and unitary. Solutions exist also in
four and higher dimensions, even and odd. In some explicit four dimensional
examples, we compute the correlation functions to the leading order in }$1/N$%
{\small \ and the critical exponents to the subleading order. We construct
also RG flows interpolating between pairs of fixed points. }

\vskip 1truecm

\vfill\eject

\section{Introduction}

\setcounter{equation}{0}

Lorentz violating quantum field theory can be useful for several purposes.
It contains non-relativistic field theory, and has applications to nuclear
physics \cite{wise1}, effective field theory \cite{physica,ergu}, critical
phenomena\/ \cite{pasquale}, and possibly high energy physics \cite{jacobson}%
. It can describe higher temperature superconductors, ferroelectric liquid
cristals, polymers and magnetic materials \cite{lifshitz,largeNlifshitz}, as
well as extensions of the Standard Model \cite{colladay} and beyond.

Moreover, Lorentz violating field theory is interesting in its own right as
a laboratory to study ideas about renormalization and learn about quantum
field theory. Recently \cite{rennolor}, it has been proved that the set of
local, unitary, renormalizable quantum field theories can be considerably
enlarged if Lorentz invariance is not assumed to hold exactly at arbitrarily
high energies. Higher space derivatives are used to improve the behavior of
propagators in Feynman diagrams. At the same time, if the vertices and
quadratic terms are arranged according to a certain ``weighted power
counting'' criterion, no higher time derivatives are generated by
renormalization, which guarantees (perturbative) unitarity. A ``weighted
scale transformation'' assigns different weights to the time and space
components of momenta and coordinates, in a way compatible with Feynman
diagrams, and is used to classify the counterterms (into weighted marginal,
weighted relevant and weighted irrelevant) and so the renormalizable models.

The weighted scale invariance is explicitly broken by the
super-renormalizable terms and dynamically broken by the running of
couplings. It is exactly recovered at the fixed points of the (weighted)
renormalization group flow. Such fixed points are worth of investigation,
because they are remarkable Lorentz violating generalizations of conformal
field theories.

In this paper we continue this investigation studying Lorentz violating
four-fermion and sigma models that are renormalizable by weighted power
counting in the large $N$ expansion, in various dimensions. Here $N$ denotes
the number of field copies. In particular, we classify the renormalizable
theories that are unitary, causal and stable. Solutions exist also in four
and higher dimensions. Some four dimensional models are studied explicitly
up to the subleading order in $1/N$.

We recall that Lorentz invariant unitary conformal field theories of scalars
and/or fermions exist in three spacetime dimensions, but not four.
Well-known examples are the three-dimensional four fermion model \cite
{parisi} and the $O(N)$ sigma model \cite{arefeva}, in the large $N$
expansion. On the other hand, in four dimensions a considerable number of
interacting conformal field theories are known, from the Bank-Zaks fixed
points \cite{cft4}, to the fixed points of supersymmetric theories \cite
{scft4}, but all of them involve gauge fields.

The models constructed in this paper contain only scalars and fermions. The
investigation of gauge theories is left to a separate paper \cite{gaugenolor}%
. Within the ordinary power counting framework, the renormalization of gauge
theories containing Lorentz violating terms has been studied in ref.s \cite
{gauge}.

For definiteness, we consider models where the $d$-dimensional spacetime
manifold $M_{d}$ is split into the product $M_{\widehat{d}}\otimes M_{%
\overline{d}}$ of two submanifolds, a $\widehat{d}$-dimensional submanifold $%
M_{\widehat{d}}$, containing time and possibly some space coordinates, and a 
$\overline{d}$-dimensional space submanifold $M_{\overline{d}}$. The $d$%
-dimensional Lorentz group $O(1,d-1)$ is broken to a residual Lorentz
symmetry $O(1,\widehat{d}-1)\otimes O(\overline{d})$. The generalization of
our arguments to the most general breaking is straightforward (see \cite
{rennolor} for details).

The paper is organized as follows. In section 2 we review the weighted power
counting criterion. In section 3 we study unitarity and stability in Lorentz
violating theories. In section 4 we study causality at the classical and
quantum levels. In sections 5 and 6 we classify the $O(N)$ sigma models, the
four-fermion models and their interacting fixed points (Lifshitz type and
Parisi type) in the large $N$ expansion. In section 7 we analyze the
consequences of the weighted scale invariance and work out restrictions on
the form of the correlations functions at the fixed points. In sections 8
and 9 we calculate the subleading corrections in a class of four-dimensional
scalar and fermion models. In section 10 we construct running models that
interpolate between pairs of fixed points. Section 11 contains our
conclusions. In Appendices A-D we calculate the bubble and triangle diagrams
in scalar and fermion models. In Appendix E we describe the calculations of
the critical exponents to the subleading order.

We use the dimensional-regularization technique, although in most formulas
we do not make it explicit. Moreover, we freely switch back and forth from
and to the Euclidean and Minkowskian frameworks, often using the same
notation.

\section{Weighted power counting}

\setcounter{equation}{0}

In this section we briefly review the weighted power counting criterion of
ref. \cite{rennolor}. Consider a scalar theory with quadratic lagrangian 
\begin{equation}
\mathcal{L}_{\hbox{free}}=\frac{1}{2}(\widehat{\partial }\varphi )^{2}+\frac{%
1}{2\Lambda _{L}^{2n-2}}(\overline{\partial }^{n}\varphi )^{2}  \label{free}
\end{equation}
(in the Euclidean framework), where $\Lambda _{L}$ is an energy scale and $n$
is an integer $\geq 1$. Up to total derivatives it is not necessary to
specify how the $2n$ derivatives $\overline{\partial }$ are contracted among
themselves. The coefficient of $(\overline{\partial }^{n}\varphi )^{2}$ must
be positive to have a positive energy in the Minkowskian framework. The
theory (\ref{free}) is invariant under the weighted rescaling 
\begin{equation}
\hat{x}\rightarrow \hat{x}\ \mathrm{e}^{-\Omega },\qquad \bar{x}\rightarrow 
\bar{x}\ \mathrm{e}^{-\Omega /n},\qquad \varphi \rightarrow \varphi \ 
\mathrm{e}^{\Omega (\hbox{\dj }/2-1)},  \label{scale}
\end{equation}
where \hbox{\dj }$=\widehat{d}+\overline{d}/n$. Indeed, each lagrangian term
scales with the factor \hbox{\dj }, compensated by the scaling factor of the
integration measure d$^{d}x$ of the action. Note that $\Lambda _{L}$ is not
rescaled.

To classify the vertices, counterterms and other quadratic terms it is
useful to assign weights to coordinates, momenta and fields as follows: 
\begin{equation}
\lbrack \widehat{\partial }]=1,\qquad [\overline{\partial }]=\frac{1}{n}%
,\qquad [\varphi ]=\frac{\hbox{\dj }-2}{2},  \label{weights}
\end{equation}
while $\Lambda _{L}$ is weightless. The interacting theory is defined as a
perturbative expansion around the free theory (\ref{free}). Strictly
renormalizable vertices have weights equal to \hbox{\dj },
super-renormalizable vertices have weights smaller than \hbox{\dj },
non-renormalizable vertices have weights greater than \hbox{\dj }. The first
condition to have renormalizability is that the $\varphi $-weight be
strictly positive, therefore \hbox{\dj } must be greater than 2.

The theory is renormalizable by weighted power counting if it contains all
vertices and quadratic terms with weights $\leq $\hbox{\dj } and only those.
This bound excludes higher-time derivative terms. The degree of divergence $%
\omega (G)$ of a Feynman diagram $G$ is bounded by the inequality 
\begin{equation}
\omega (G)\leq \hbox{\dj }-E_{s}\frac{\hbox{\dj }-2}{2},  \label{abbound}
\end{equation}
where $E_{s}$ is the number of external scalar legs. Formula (\ref{abbound})
ensures that the counterterm has a weight not larger than \dj , therefore it
can be subtracted renormalizing the fields and couplings of the lagrangian,
and no new vertex needs to be introduced.

Strictly renormalizable theories are called ``homogeneous''. The propagator
of homogeneous theories coincides with the one of (\ref{free}). The bound %
\hbox{\dj }$>2$ ensures that in homogeneous theories the Feynman diagrams do
not have infrared divergences at non-exceptional external momenta.

The RG flow measures how correlation functions depend on the overall
weighted rescaling factor. When some couplings run, the weighted scale
transformation is anomalous. The ``weighted trace anomaly'' is parametrized
by the beta functions. At the fixed points of the RG flow the weighted scale
invariance is recovered as an exact symmetry.

In the ordinary perturbative framework, stable renormalizable interacting
theories exist for \hbox{\dj }$\leq 4$. Unstable renormalizable theories,
such as the $\varphi ^{3}$ models, exist for \hbox{\dj }$\leq 6$. The
simplest examples of stable, homogeneous theories are the $\varphi ^{4}$, %
\hbox{\dj }$=4$ models 
\begin{equation}
\mathcal{L}_{\hbox{\dj }=4}=\frac{1}{2}(\widehat{\partial }\varphi )^{2}+%
\frac{1}{2\Lambda _{L}^{2(n-1)}}(\overline{\partial }^{n}\varphi )^{2}+\frac{%
\lambda }{4!\Lambda _{L}^{d-4}}\varphi ^{4}  \label{scal}
\end{equation}
and the $\varphi ^{6}$, \hbox{\dj }$=3$ models 
\begin{equation}
\mathcal{L}_{\hbox{\dj }=3}=\frac{1}{2}(\widehat{\partial }\varphi )^{2}+%
\frac{1}{2\Lambda _{L}^{2(n-1)}}(\overline{\partial }^{n}\varphi )^{2}+\frac{%
1}{4!\Lambda _{L}^{2(n-1)}}\sum_{\alpha }\lambda _{\alpha }\left[ {\overline{%
\partial }}^{n}{\varphi }^{4}\right] _{\alpha }+\frac{\lambda _{6}}{%
6!\Lambda _{L}^{2(n-1)}}\varphi ^{6},  \label{even}
\end{equation}
where $\left[ {\overline{\partial }}^{n}{\varphi }^{4}\right] _{\alpha }$
denotes a basis of inequivalent terms constructed with $n$ derivatives ${%
\overline{\partial }}${\ acting on four }${\varphi }$'s (because of $O(%
\overline{d})$-invariance, these exist no such terms if $n$ is odd).

The considerations just recalled are easily generalized to fermions. The
weight of a fermion field is (\hbox{\dj }$-1)/2$, so renormalizability
demands \hbox{\dj }$>1$. Again, this bound ensures also that the Feynman
diagrams are free of infrared divergences at non-exceptional external
momenta in homogeneous theories. Renormalizable theories are those that
contain all vertices and quadratic terms with weight not larger than \dj\
and only those. Nontrivial stable renormalizable theories containing only
fermions exist for \hbox{\dj }$\leq 2$. The simplest homogeneous examples
are the \dj $=2$, four-fermion models 
\begin{equation}
\mathcal{L}_{\hbox{\dj }=2}=\overline{\psi }\left( \widehat{\partial }\!\!\!%
\slash+\frac{{\overline{\partial }\!\!\!\slash\,}^{n}}{\Lambda _{L}^{n-1}}%
\right) \psi -\frac{\lambda ^{2}}{2\Lambda _{L}^{d-2}}\left( \overline{\psi }%
\psi \right) ^{2}.  \label{ferm}
\end{equation}

Stable coupled scalar and fermion theories exist for \hbox{\dj }$\leq 4$.
Formula (\ref{abbound}) becomes 
\[
\omega (G)\leq \hbox{\dj }-E_{s}\frac{\hbox{\dj }-2}{2}-E_{f}\frac{%
\hbox{\dj 
}-1}{2}, 
\]
where $E_{f}$ is the number of external fermionic legs.

Non-homogeneous renormalizable theories contain also super-renormalizable
quadratic terms and vertices. For convenience, the coefficient of each
vertex is arranged as the product of three factors: $a$) a suitable power of
a mass scale $M$ of weight 1, to match the total weight; $b$) a suitable
power of $\Lambda _{L}$, to match the dimensionality; $c$) a dimensionless
weightless coupling $\lambda _{i}$. In such a way, super-renormalizable
vertices are multiplied by positive powers of $M$, strictly-renormalizable
vertices are multiplied by $M$-independent coefficients, while
non-renormalizable vertices are multiplied by negative powers of $M$.

For example, in $2<$\hbox{\dj }$<4$ the $\varphi ^{4}$-model is
super-renormalizable, with lagrangian 
\begin{equation}
\mathcal{L}_{\hbox{\dj }<4}=\frac{1}{2}(\widehat{\partial }\varphi
)^{2}+\sum_{k=0}^{n}\left. ^{\prime }\right. \frac{\lambda _{k}M^{2(1-k/n)}}{%
2\Lambda _{L}^{2k(n-1)/n}}(\overline{\partial }^{k}\varphi )^{2}+\frac{%
\lambda M^{4-\hbox{\textit{\dj }}}}{4!\Lambda _{L}^{\overline{d}(1-1/n)}}%
\varphi ^{4},  \label{lif}
\end{equation}
with $\lambda _{n}=1$, and the primed sum is restricted to the $k$'s such
that $2(1-k/n)$ are integer multiples of $4-$\dj . Other examples of
super-renormalizable theories are the four-fermion models in $1<$\hbox{\dj }$%
<2$ with lagrangian 
\begin{equation}
\mathcal{L}_{\hbox{\dj }<2}=\overline{\psi }\left( \widehat{\partial }\!\!\!%
\slash+\sum_{k=0}^{n}\left. ^{\prime }\right. \frac{\lambda _{k}M^{1-k/n}{%
\overline{\partial }\!\!\!\slash\,}^{k}}{\Lambda _{L}^{k(n-1)/n}}\right)
\psi -\frac{\lambda ^{2}M^{2-\hbox{\dj }}}{2\Lambda _{L}^{\overline{d}%
(1-1/n)}}\left( \overline{\psi }\psi \right) ^{2},  \label{par}
\end{equation}
where now the primed sum is restricted to the $k$'s such that $1-k/n$ are
integer multiples of $2-$\dj .

The RG flow that we consider in this paper is more precisely the ``weighted
RG flow'', defined by the weights of the fields and couplings, rather than
by their dimensionalities. In particular, the infrared limit is the limit
where both $M$ and the RG scale $\mu $ tend to infinity, while $\Lambda _{L}$
is kept fixed. Analogously, the ultraviolet limit is defined as the limit
where both $M$ and $\mu $ tend to zero, at fixed $\Lambda _{L}$. As a
consequence, the fixed points of the weighted RG flow do depend on $\Lambda
_{L}$.

\section{K\"{a}llen-Lehmann representation and unitarity}

\setcounter{equation}{0}

In this section we study unitarity and stability in Lorentz violating
quantum field theory, generalizing the usual notions.

Let $\left| n\right\rangle $ be a complete set of eigenstates of the
momentum, with eigenvalues $k_{n}$. Consider the sum 
\begin{equation}
\sum_{n}\delta ^{(d)}(k-k_{n})\left| \left\langle 0\right| \varphi (0)\left|
n\right\rangle \right| ^{2},  \label{u1}
\end{equation}
where $\varphi $ is any (real) scalar field, elementary or composite. By $%
O(1,\widehat{d}-1)\otimes O(\overline{d})$ invariance this sum can depend
only on $\widehat{k}^{2}$, $\overline{k}^{2}$ and, for $\widehat{k}^{2}>0$,
on $\theta (k_{0})$. Moreover, by stability it is zero for $\widehat{k}^{2}<0
$ and for $k_{0}\leq 0$, because $\widehat{k}_{n}^{2}\geq 0$, $k_{0}>0$ for
every contributing $n$ (we assume that $\varphi $ has no vacuum expectation
value, so $\left\langle 0\right| \varphi (0)\left| 0\right\rangle =0$). If
we write 
\begin{equation}
\sum_{n}\delta ^{(d)}(k-k_{n})\left| \left\langle 0\right| \varphi (0)\left|
n\right\rangle \right| ^{2}=\theta (k_{0})\rho (\widehat{k}^{2},\overline{k}%
^{2})  \label{u2}
\end{equation}
unitarity tells us that the spectral function $\rho (\widehat{k}^{2},%
\overline{k}^{2})$ is real and positive. The two-point function of $\varphi $
can be written as 
\begin{eqnarray}
\left\langle 0\right| \varphi (x)\varphi (0)\left| 0\right\rangle 
&=&\sum_{n}\mathrm{e}^{-ik_{n}\cdot x}\left| \left\langle 0\right| \varphi
(0)\left| n\right\rangle \right| ^{2}=\int \frac{\mathrm{d}^{d}k}{(2\pi )^{d}%
}\mathrm{e}^{-ik\cdot x}\theta (k_{0})\rho (\widehat{k}^{2},\overline{k}^{2})
\nonumber \\
&=&\int_{0}^{\infty }\mathrm{d}s\int \frac{\mathrm{d}^{d}k}{(2\pi )^{d}}%
\mathrm{e}^{-ik\cdot x}\theta (k_{0})\delta (\widehat{k}^{2}-s)\rho (s,%
\overline{k}^{2}).  \label{u3}
\end{eqnarray}
Using 
\[
\int \frac{\mathrm{d}^{\widehat{d}}\widehat{k}}{(2\pi )^{\widehat{d}}}%
\mathrm{e}^{-i\widehat{k}\cdot \widehat{x}}\left[ \theta (x_{0})\theta
(k_{0})+\theta (-x_{0})\theta (-k_{0})\right] \delta (\widehat{k}%
^{2}-s)=\int \frac{\mathrm{d}^{\widehat{d}}\widehat{k}}{(2\pi )^{\widehat{d}}%
}\frac{i\mathrm{e}^{-i\widehat{k}\cdot \widehat{x}}}{\widehat{k}%
^{2}-s+i\varepsilon },
\]
the time-ordered correlation function reads 
\begin{equation}
\Delta (x)\equiv \left\langle 0\right| T\varphi (x)\varphi (0)\left|
0\right\rangle =\int_{0}^{\infty }\mathrm{d}s\int \frac{\mathrm{d}^{d}k}{%
(2\pi )^{d}}\frac{i\mathrm{e}^{-ik\cdot x}\rho (s,\overline{k}^{2})}{%
\widehat{k}^{2}-s+i\varepsilon }.  \label{u4}
\end{equation}
The spectral function coincides with the imaginary part of $i/\pi $ times
the Fourier transform of $\Delta (x)$: 
\begin{equation}
\hbox{Im}\left[ \frac{i}{\pi }\left\langle \widetilde{\varphi }(-k)%
\widetilde{\varphi }(k)\right\rangle \right] =\rho (\widehat{k}^{2},%
\overline{k}^{2})\geq 0.  \label{ineq}
\end{equation}
Therefore, the Fourier transform $\left\langle \widetilde{\varphi }(-k)%
\widetilde{\varphi }(k)\right\rangle $ of the T-ordered two-point function
has the spectral representation 
\begin{equation}
i\left\langle \widetilde{\varphi }(-k)\widetilde{\varphi }(k)\right\rangle
=\int_{0}^{\infty }\frac{\rho (s,\overline{k}^{2})\mathrm{d}s}{s-\widehat{k}%
^{2}-i\varepsilon }.  \label{spec}
\end{equation}
Further, defining 
\begin{equation}
\Delta ^{\pm }(x)\equiv \int \frac{\mathrm{d}^{d}k}{(2\pi )^{d}}\mathrm{e}%
^{-ik\cdot x}\theta (\pm k_{0})\rho (\widehat{k}^{2},\overline{k}^{2}),
\label{u5}
\end{equation}
we have immediately 
\begin{equation}
\Delta (x)=\theta (x_{0})\Delta ^{+}(x)+\theta (-x_{0})\Delta ^{-}(x)
\label{u6}
\end{equation}
and the relations 
\begin{equation}
\Delta ^{\mp }(-x)=\Delta ^{\pm }(x),\qquad \Delta ^{\pm *}(x)=\Delta ^{\mp
}(x),\qquad \Delta ^{*}(x)=\theta (x_{0})\Delta ^{-}(x)+\theta
(-x_{0})\Delta ^{+}(x).  \label{u7}
\end{equation}
The ``dressed'' propagators $\Delta ^{\pm }(x)$ can be used to define
cutting Feynman rules and cutting diagrams, as usual. The cutting method
allows us to calculate the imaginary parts of diagrams, thanks to the
unitarity equation $iT-iT^{\dagger }=-T^{\dagger }T$, where $S=1+iT$ is the $%
S$-matrix. The equation is graphically illustrated in Fig. 1 \cite
{diagrammar}.

\begin{figure}[tbp]
\centerline{\includegraphics[width=2.3in,height=.5in]{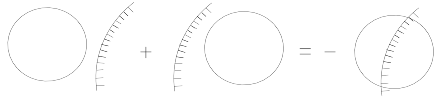}}
\caption{unitarity}
\end{figure}

For example, the tree-level scalar Minkowskian propagator reads 
\begin{equation}
\frac{i}{\widehat{k}^{2}-f(\overline{k}^{2})+i\varepsilon }  \label{propa}
\end{equation}
for some positive function $f$. Then (\ref{ineq}) gives 
\[
\rho (\widehat{k}^{2},\overline{k}^{2})=\delta \left( \widehat{k}^{2}-f(%
\overline{k}^{2})\right) . 
\]
The cutting propagators read in momentum space 
\[
(2\pi )\widetilde{\Delta }^{\pm }(k)=(2\pi )\theta (\pm k_{0})\delta \left( 
\widehat{k}^{2}-f(\overline{k}^{2})\right) =\frac{\pi }{\sqrt{\widehat{%
\mathbf{k}}^{2}+f(\overline{k}^{2})}}\delta \left( k_{0}\mp \sqrt{\widehat{%
\mathbf{k}}^{2}+f(\overline{k}^{2})}\right) , 
\]
depending on the orientation of the energy flow with respect to the cut.

\paragraph{Unitarity bounds}

The spectral function $\rho (s,\overline{k}^{2})$ must be regular. In
particular, it must be integrable at $s=0$ for every value of $\overline{k}%
^{2}$ and grow at most as fast as a polynomial in $s$ when $s$ is large. In
this paper we deal with spectral functions that are manifestly regular for $%
\overline{k}^{2}\neq 0$ and behave correctly for $s$ large, but the behavior
of $\rho (s,0)$ for $s$ small needs to be carefully checked. In weighted
scale invariant theories we can assume that $\rho (s,0)$ has a power-like
behavior. Then the integrability of (\ref{spec}) demands 
\begin{equation}
\rho (s,0)\sim \frac{1}{s^{a}}\qquad \hbox{with }a<1\qquad \hbox{for }%
s\rightarrow 0.  \label{a<1}
\end{equation}
The condition $a<1$ is a powerful requirement to constrain the range of
values of \hbox{\dj } for which unitarity holds.

We recall that the RG fixed points of Lorentz invariant local quantum field
theories containing fields of spin $\leq 1$ are also conformal field
theories. There, the unitarity bound (\ref{a<1}) ensures that the
dimensionality of a scalar (primary) field $\varphi $ is not smaller than
one. We emphasize that we do not need conformal invariance to derive the
unitarity bound (\ref{a<1}). As we see, not even Lorentz invariance is
necessary.

\paragraph{Stability}

In perturbative quantum field theory stability, as well as unitarity, must
be checked at the leading order. In general, this means at the tree level,
but in the models of this paper one field, $\sigma $, has a dynamically
generated propagator. At the leading order of the large $N$ expansion the $%
\sigma $-propagator is determined by a one-loop diagram (see fig. 2).

The $\sigma $ spectral function $\rho _{\sigma }(\widehat{k}^{2},\overline{k}%
^{2})$ must vanish for $\widehat{k}^{2}<0$ and be non-negative for $\widehat{%
k}^{2}>0$. We can prove in complete generality that these requirements are
automatically fulfilled. Call $\sigma _{\mathrm{M}}$ and $iT$ the
Minkowskian $\sigma $ field and bubble diagram, respectively. The unitarity
relation of fig. 1 tells us that the imaginary part of $T$ can be calculated
using the cutting technique, it is convergent and non-negative. Therefore,
at the leading order in $1/N$ we have $\left\langle \widetilde{\sigma }_{%
\mathrm{M}}(-k)\widetilde{\sigma }_{\mathrm{M}}(k)\right\rangle =i/T(k)$, so
the $\sigma $ spectral function reads 
\[
\rho _{\sigma }(\widehat{k}^{2},\overline{k}^{2})\equiv \hbox{Im}\left[ 
\frac{i}{\pi }\left\langle \widetilde{\sigma }_{\mathrm{M}}(-k)\widetilde{%
\sigma }_{\mathrm{M}}(k)\right\rangle \right] =\frac{\hbox{Im}T(k)}{\pi
|T(k)|^{2}}\geq 0 
\]
and is necessarily non-negative. Moreover, writing the cutting diagram of $%
\hbox{Im}T(k)$ explicitly, it is straightforward to check that $\rho
_{\sigma }(\widehat{k}^{2},\overline{k}^{2})$ vanishes for $\widehat{k}%
^{2}<0 $, both in our scalar and fermion models. The spectral representation
(\ref{spec}), the other formulas from (\ref{u1}) to (\ref{u7}) and the
cutting rules apply to the $\sigma $ field with obvious adjustments.

Because of stability, the bosonic sectors of the classical action and of the
generating functional $\Gamma $ of one-particle irreducible diagrams must be
positive definite in the Euclidean framework. No general argument guarantees
the positivity of the $\sigma $-sector of $\Gamma $ (counterexamples are
easy to construct), therefore this aspect needs to be investigated in detail.

\paragraph{Regularity of the $\sigma $ propagator}

To verify the consistency of our theories, it is necessary, in addition, to
check that the $\sigma $ propagator $P_{w}(\widehat{k},\overline{k})$ be
regular everywhere. Here $-w$ denotes its weight. In particular, in the
``ultraviolet'' limits $\widehat{k}\rightarrow \infty $ and $\overline{k}%
\rightarrow \infty $, $P_{w}$ must behave as 
\begin{equation}
P_{w}(\widehat{k},\overline{k})\sim \frac{1}{|\widehat{k}|^{w}},\qquad P_{w}(%
\widehat{k},\overline{k})\sim \frac{1}{|\overline{k}|^{nw}},  \label{uvbeha}
\end{equation}
respectively, to be consistent with the weighted power counting. For
example, the propagator of (\ref{free}), 
\[
\frac{1}{\widehat{k}^{2}+\frac{(\overline{k}^{2})^{n}}{\Lambda _{L}^{2(n-1)}}%
},
\]
is regular of weight $-$2, but a propagator of the form 
\[
\frac{\Lambda _{L}^{n-1}}{|\widehat{k}||\overline{k}|^{n}}
\]
is not regular, and could generate spurious ultraviolet sub-divergences in
Feynman diagrams when $\widehat{k}$ tends to infinity at $\overline{k}$
fixed, or viceversa (see ref. \cite{rennolor} for details). In some of our
models (e.g. the one studied in section 8) regularity can be proved
straightforwardly, in other models (e.g. the one studied in section 9)
regularity is not fulfilled and the absence of spurious divergences has to
be proved by direct analysis.

\section{Causality}

\setcounter{equation}{0}

In this section we investigate causality at the classical and quantum
levels. We work of course in the Minkowskian framework.

\paragraph{Classical theory}

Consider the scalar field theory 
\[
L=\frac{1}{2}(\widehat{\partial }\varphi )^{2}-\frac{1}{2}\varphi f(-%
\overline{\partial }^{2})\varphi +J\varphi 
\]
coupled with an external source $J$, where $f$ is a positive polynomial
function. The field equations 
\begin{equation}
\left[ -\widehat{\partial }^{2}-f(-\overline{\partial }^{2})\right] \varphi
(x)=J(x)  \label{efqs}
\end{equation}
are solved as 
\begin{equation}
\varphi (x)=\int G_{\hbox{ret}}(x-x^{\prime })\varphi (x^{\prime })\mathrm{d}%
^{d}x^{\prime }  \label{sola}
\end{equation}
where $G_{\hbox{ret}}(x-x^{\prime })$ is the retarded Green function.
Decomposing $\widehat{k}$ as $(k_{0},\widehat{\mathbf{k}})$, we have 
\[
G_{\matrix{ \text{ret (adv)}}}(x)=\int \frac{\mathrm{d}^{d}k}{(2\pi )^{d}}%
\frac{i\mathrm{e}^{-ik\cdot x}}{(k_{0}\pm i\varepsilon )^{2}-\widehat{%
\mathbf{k}}^{2}-f(\overline{k}^{2})}. 
\]
The retarded (advanced) Green function vanishes for negative (positive) time
intervals, for an arbitrary function $f(\overline{k}^{2})$. Thus the
solution (\ref{sola}) of the field equations (\ref{efqs}) is determined by
the sole knowledge of the source $J$ in the present and in the past, which
ensures causality at the classical level.

\paragraph{Quantum theory}

Bogoliubov's definition of causality reads \cite{bogoliubov} 
\begin{equation}
\frac{\delta ^{2}S}{\delta g(x_{i})\delta g(x_{j})}S^{\dagger }+\frac{\delta
S}{\delta g(x_{i})}\frac{\delta S^{\dagger }}{\delta g(x_{j})}=0\qquad %
\hbox{if }x_{i0}<x_{j0},  \label{uno}
\end{equation}
where $S$ denotes the $S$-matrix and $g(x)$ is a coupling constant, made
into a function of spacetime. Formula (\ref{uno}) admits a more general
single-diagram version \cite{diagrammar}, which reads 
\begin{equation}
\sum_{\matrix{\text{underlinings except }x_{i}}}F(x_{1},\ldots
,x_{n})=0\qquad \hbox{if }x_{i0}<x_{j0}.  \label{due}
\end{equation}
Here $F(x_{1},\ldots ,x_{n})$ denotes a diagram in coordinate space with
vertices in $x_{1},\ldots ,x_{n}$. No integral over the positions of the
vertices is understood. The propagator $\Delta (x_{k}-x_{l})$ connects two
non-underlined points $x_{k}$ and $x_{l}$, while $\Delta ^{-}(x_{k}-x_{l})$
connects a non-underlined point $x_{k}$ with an underlined point $\underline{%
x_{l}}$ , $\Delta ^{+}(x_{k}-x_{l})$ connects $\underline{x_{k}}$ with $%
x_{l} $ and $\Delta ^{*}(x_{k}-x_{l})$ connects $\underline{x_{k}}$ with $%
\underline{x_{l}}$. Finally, every underlined vertex carries an extra minus
sign.

The proof of (\ref{due}) is done as follows. Call $x_{k}$ the vertex with
the largest time component $x_{k0}$. By assumption, $x_{k}$ is not $x_{i}$.
Hence the sum (\ref{due}) can be rearranged into the sum over pairs of
diagrams differing only by $x_{k}$ being underlined or not. We want to show
that the diagrams of each pair sum to zero. Indeed, they differ by an
overall minus sign, because of the $x_{k}$ underlining, and have exactly the
same propagators. To see this, let $x_{m}$ denote any vertex connected with $%
x_{k}$. Since $x_{k0}>x_{m0}$ the identities 
\begin{equation}
\Delta (x_{k}-x_{m})=\Delta ^{+}(x_{k}-x_{m}),\qquad \Delta
^{*}(x_{k}-x_{m})=\Delta ^{-}(x_{k}-x_{m}),  \label{ide}
\end{equation}
hold. Therefore, within each pair it does not matter whether $x_{k}$ is
underlined or not, apart from the relative minus sign. We conclude that the
sum (\ref{due}) vanishes identically.

Finally, formula (\ref{uno}) is derived multiplying (\ref{due}) by
appropriate source functions, integrating over all points but $x_{i}$ and $%
x_{j}$, and summing over all diagrams. The first term of (\ref{uno})
collects the diagrams where neither $x_{i}$ nor $x_{j}$ are underlined,
while the second term of (\ref{uno}) collects the diagrams where $x_{i}$ is
not underlined, but $x_{j}$ is.

Our Lorentz violating theories are automatically causal in Bogoliubov's
sense. This is obvious for the models of \cite{rennolor}, which involve no
large $N$ expansion, but true also in the models studied here, because the $%
\sigma $ propagator satisfies the cutting rules and formulas (\ref{u4})-(\ref
{u7}).

On the other hand, the usual operator relation 
\begin{equation}
\lbrack \varphi (x),\varphi (y)]=0,\qquad \hbox{ for }x-y%
\hbox{ =
spacelike,}  \label{oldcau}
\end{equation}
which is a consequence of (\ref{uno}), unitarity and Lorentz invariance \cite
{bogoliubov}, is meaningless in our theories. Yet, (\ref{oldcau}) is not
necessary to have causality and unitarity.

\section{Lifshitz type models}

\setcounter{equation}{0}

In this section we construct scalar models that have nontrivial interacting
fixed points in the large $N$ expansion and study those fixed points. We
classify the unitary, stable models and single out the four dimensional
cases.

Consider the $O(N)$ sigma model 
\begin{equation}
\mathcal{L}_{\eta }=\frac{1}{2}\sum_{i=1}^{N}\left[ (\widehat{\partial }%
\varphi _{i})^{2}+\frac{1}{\Lambda _{L}^{2n-2}}(\overline{\partial }%
^{n}\varphi _{i})^{2}\right] +\frac{1}{2}i\lambda \sigma \left(
\sum_{i=1}^{N}\varphi _{i}^{2}-\eta ^{2}\right) ,  \label{sigma}
\end{equation}
in the Euclidean framework, where $\eta $ is a positive constant and $\sigma 
$ is a field. Integrating over $\sigma $ constrains the scalar field to live
on a sphere of radius $\eta $. The redundant parameter $\lambda $ is
introduced for convenience. The large $N$ expansion is defined as the
expansion in $1/N$, where $N$ is sent to infinity keeping $\lambda ^{2}N$
finite.

The field $\sigma $ does not have a kinetic term. In the large $N$ expansion
the missing $\sigma $ propagator is generated dynamically. Precisely, it is
equal to minus the reciprocal of the scalar bubble of Fig. 2, whose form is (%
\ref{sbubble}) (in the case $\widehat{d}=3$, $\overline{d}=1$, $n=2$). Every
other diagram gives a subleading contribution.

The constant $\eta $ has a positive weight (for \hbox{\dj }$>2$) and a
positive dimensionality, therefore the UV fixed point is the same theory
with $\eta =0$: 
\begin{equation}
\mathcal{L}_{C}=\frac{1}{2}\sum_{i=1}^{N}\left[ (\widehat{\partial }\varphi
_{i})^{2}+\frac{1}{\Lambda _{L}^{2n-2}}(\overline{\partial }^{n}\varphi
_{i})^{2}+i\lambda \sigma \varphi _{i}^{2}\right] .  \label{fix}
\end{equation}
Moreover, the term $-i\lambda \sigma \eta ^{2}/2$ does not contribute to any
non-trivial one-particle irreducible diagram, therefore the generating
functionals $\Gamma _{\eta }$ and $\Gamma _{C}$ of (\ref{sigma}) and (\ref
{fix}) differ exactly by that term. For the same reason, once we prove the
renormalizability of (\ref{fix}) we prove also the renormalizability of (\ref
{sigma}).

The theory (\ref{fix}) coincides with the Lifshitz fixed point of the $O(N)$ 
$\varphi ^{4}$-theory (\ref{lif}), which in the cases $n=2$, \hbox{\dj }$<4$
has been studied in \cite{lifshitz,largeNlifshitz}. Here we are interested
in more general situations, so we allow \hbox{\dj } to be greater than 4 and
keep $n$ generic.

In the Minkowskian framework the lagrangian of the fixed point reads 
\begin{equation}
\mathcal{L}_{\mathrm{M}}=\frac{1}{2}\sum_{i=1}^{N}\left[ (\widehat{\partial }%
_{\mathrm{M}}\varphi _{i})^{2}-\frac{1}{\Lambda _{L}^{2n-2}}(\overline{%
\partial }^{n}\varphi _{i})^{2}+\lambda \sigma _{\mathrm{M}}\varphi
_{i}^{2}\right] ,  \label{minko}
\end{equation}
which is Hermitian if $\sigma _{\mathrm{M}}=-i\sigma $ is real.

Let us describe some basic properties of the theory (\ref{fix}), assuming
for the moment that it is renormalizable as it stands. The conditions for
renormalizability are worked out below. Since a renormalization constant in
front of the vertex $\sigma \varphi ^{2}$ can be interpreted as the $\sigma $
wave function renormalization constant (instead of the $\lambda $
renormalization constant), the theory has no true coupling, although it is
interacting. As a consequence, there is no running of couplings, so the
model (\ref{fix}) is exactly invariant under the weighted scale
transformation 
\begin{equation}
\hat{x}\rightarrow \hat{x}\ \mathrm{e}^{-\Omega },\qquad \bar{x}\rightarrow 
\bar{x}\ \mathrm{e}^{-\Omega /n},\qquad \varphi _{i}\rightarrow \varphi
_{i}\ \mathrm{e}^{\Omega (\hbox{\dj }/2-1)},\qquad \sigma \rightarrow \sigma
\ \mathrm{e}^{2\Omega },\qquad \mu \rightarrow \mu \ \mathrm{e}^{\Omega },
\label{scale2}
\end{equation}
also at the quantum level, once we include the rescaling of the RG scale $%
\mu $.

\begin{figure}[tbp]
\centerline{\includegraphics[width=1.5in,height=.6in]{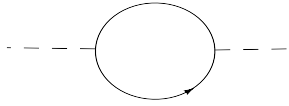}}
\caption{Bubble diagram that gives the $\sigma $-propagator}
\end{figure}
Next, since the weight of $\varphi $ must be positive, we have \hbox{\dj }$%
>2 $, which ensures also, from section 2, that Feynman diagrams do not
contain IR divergences at non-exceptional external momenta. We have observed
in section 4 that causality is always guaranteed. Nevertheless, unitarity,
renormalizability and stability are not obvious and put other restrictions
on the allowed values of \hbox{\dj }.

\paragraph{Unitarity}

In section 3 we have proved in complete generality that the spectral
function $\rho _{\sigma }(\widehat{p}^{2},\overline{p}^{2})$ of the $\sigma
_{\mathrm{M}}$ field in the Minkowskian framework is non-negative and
vanishes for $\widehat{p}^{2}<0$. It remains to study the unitary bound (\ref
{a<1}). Since (\ref{fix}) is a homogeneous theory, $\rho _{\sigma }(s,0)$ is
a homogeneous function. Since $s$ stands for $\widehat{p}^{2}$, the weight
of $s$ is equal to $2$. On the other hand, the weight of $\rho _{\sigma
}(s,0)$ is equal to minus the weight of the scalar bubble. We conclude that 
\[
\rho _{\sigma }(s,0)\sim \frac{1}{s^{\hbox{\dj }/2-2}} 
\]
for $s\sim 0$, so the unitarity bound gives \hbox{\dj }$<6$.

\paragraph{Renormalizability}

Now we study the counterterms. Since we use the dimensional-regularization
technique, no power-like divergences are generated. The theory contains no
weightful parameter (recall that $\Lambda _{L}$ is weightless), therefore
every counterterm must be of weight \dj . To have exact weighted scale
invariance at the quantum level, (\ref{fix}) must be renormalizable as it
stands, so no new vertex should be generated by renormalization. Indeed, a
new vertex would be multiplied, in general, by a running coupling constant,
so (\ref{fix}) would no longer be a fixed point of the RG flow. We need to
classify all counterterms compatible with locality and weighted power
counting. Consider first the counterterms of the form 
\begin{equation}
\sigma \overline{\partial }^{q}\sigma .  \label{sissi}
\end{equation}
We must exclude them for every even non-negative integer $q$. The weight of (%
\ref{sissi}) is $4+q/n$, which by homogeneity must be equal to \hbox{\dj }.
Therefore we have 
\[
q=n(\hbox{\dj }-4)=\overline{d}(1-n)+n(d-4).
\]
Now, this $q$ is negative for \hbox{\dj }$<4$, and zero for \hbox{\dj }$=4$.
Thus \hbox{\dj }$=4$ must be excluded. Moreover, in the range $4<$\hbox{\dj }%
$<6$ we have to exclude every case where $q$ is even. This leaves only two
situations: $i$) both $n$ and $d$ are odd; $ii$) $n$ is even and $\overline{d%
}$ is odd.

Next, consider 
\begin{equation}
\lbrack \widehat{\partial }^{p}\overline{\partial }^{q}]\sigma ^{m},
\label{yu}
\end{equation}
where the square bracket is used as a symbolic notation to denote any action
of the derivatives on the $\sigma $ fields. Equating the weight of (\ref{yu}%
) to \hbox{\dj } we have 
\[
p+\frac{q}{n}=\hbox{\dj }-2m\hbox{\textit{,}}
\]
and since \hbox{\dj }$<6$ either $p$ or $q$ is necessarily negative for
every $m\geq 3$. Therefore, the absence of (\ref{yu}) does not impose new
restrictions on \hbox{\dj }.

There exists only one term of weight \hbox{\dj } containing both $\sigma $'s
and $\varphi $'s, which is $\sigma \varphi ^{2}$. Thus it remains to
consider those counterterms that contain only $\varphi $ fields. Any time a
counterterm factorizes a $\sum_{i=1}^{N}\varphi _{i}^{2}$ it is proportional
to the $\sigma $ field equation, so it can be removed redefining the field $%
\sigma $. Consider the counterterms containing $2m$ $\varphi $'s and a
certain number $q$ of derivatives $\overline{\partial }$%
\begin{equation}
\lbrack \overline{\partial }^{q}]\varphi _{i_{1}}\varphi _{i_{1}}\cdots
\varphi _{i_{m}}\varphi _{i_{m}}.  \label{conte}
\end{equation}
We do not need to consider derivatives $\widehat{\partial }$ since there
would be at least two of them and then the counterterm would necessarily
contain just two $\varphi $'s and no $\overline{\partial }$, otherwise its
weight would exceed \hbox{\dj }. It is easy to prove that if $q$ is smaller
than $2m$ the counterterm (\ref{conte}) is always proportional to $%
\sum_{i=1}^{N}\varphi _{i}^{2}$, up to total derivatives. Instead, if $q$ is
at least $2m$, then there exists an arrangement, namely 
\[
\prod_{k=1}^{m}(\overline{\partial }\varphi _{i_{k}})^{2},
\]
that is not proportional to $\sum_{i=1}^{N}\varphi _{i}^{2}$. Thus, writing $%
q=2m+\Delta q$, the counterterm (\ref{conte}) should be forbidden for every
non-negative even integer $\Delta q$. The weight condition gives 
\begin{equation}
\Delta q=2m(n-1)-n(m-1)d+\overline{d}(m-1)(n-1).  \label{sho}
\end{equation}
This formula implies that the counterterm (\ref{conte}) is automatically
forbidden, or reabsorbable into the $\sigma $ field equation, when 
\begin{equation}
\hbox{\dj }>\frac{2m(n-1)}{n(m-1)},  \label{iner}
\end{equation}
because then $\Delta q$ is negative. On the other hand, (\ref{sho}) also
shows that if $m$ is odd $\Delta q$ is certainly even. In particular, the
counterterm (\ref{conte}) with $m=3$ is forbidden or reabsorbable into the $%
\sigma $ field equation if and only if (\ref{iner}) holds, namely 
\begin{equation}
\hbox{\dj }>3-\frac{3}{n}.  \label{iner2}
\end{equation}
Luckily, (\ref{iner2}) excludes also every counterterm (\ref{conte}) with $%
m>3$, because it implies (\ref{iner}). Yet, it remains to exclude the
counterterm with $m=2$, which requires either 
\[
\hbox{\dj }>4-\frac{4}{n},
\]
or one of the situations $i$)-$ii$) mentioned before.

Summarizing, we have unitary renormalizable models in the following three
situations: 
\begin{eqnarray}
1.\qquad \hbox{\dj } &>&2\hbox{,\qquad \qquad }3-\frac{3}{n}<%
\hbox{\textit{\dj }}\leq 4-\frac{4}{n},\qquad \hbox{and either }i\hbox{) or }%
ii\hbox{);}  \label{tre} \\
2.\qquad \hbox{\dj } &>&2\hbox{,\qquad\qquad  }4-\frac{4}{n}<%
\hbox{\textit{\dj }}<4\hbox{;}  \label{treprimo} \\
3.\qquad 4 &<&\hbox{\dj }<6,\qquad \hbox{and either }i\hbox{) or }ii\hbox{).}
\label{tresecondo}
\end{eqnarray}
We recall that $i$) means that both $n$ and $d$ are odd, while $ii$) means
that $n$ is even and $\overline{d}$ is odd.

\paragraph{Stability}

At the leading order in the $1/N$ expansion stability can be checked
verifying that the $\sigma $-quadratic contribution to the generating
functional $\Gamma $ is positive definite in the Euclidean framework. For %
\hbox{\dj }$<4$ the positivity of the $\sigma $ two-point function is
guaranteed. Indeed, the $\sigma $ bubble (see (\ref{sbubble})) is a
convergent integral of a negative definite integrand (because of the factor $%
-\lambda ^{2}N/2$), so it is negative. The $\sigma $ propagator is minus the
reciprocal of the $\sigma $ bubble, so it is positive. This argument does
not apply for $4<$\hbox{\dj }$<6$. Indeed, in that range the bubble diagram
is formally divergent and stability has to be checked explicitly.

\paragraph{Regularity}

For analogous reasons, the $\sigma $ propagator is manifestly regular for %
\hbox{\dj }$<4$. Indeed, since the bubble integral is convergent and its
integrand is negative definite, setting $\overline{k}=0$ or $\widehat{k}=0$
gives precisely the power-like behaviors (\ref{uvbeha}) with $w=$\hbox{\dj
}$-4$. For $4<$\hbox{\dj }$<6$ regularity has to be checked explicitly case
by case.

\bigskip

The models that satisfy (\ref{tre}) and (\ref{treprimo}) are guaranteed to
be unitary, stable and regular. Let us list the four dimensional solutions.
Clearly, $\widehat{d}$ must be $1$, $2$ or $3$. For $\widehat{d}=1$, $2$ the
unique solution is $n=2$, with \hbox{\dj }$=5/2$, $3$, respectively, while
for $\widehat{d}=3$, $n$ can be an arbitrary even number greater than one,
or equal to 3. The $n=2$, \hbox{\dj }$=7/2$ four dimensional model is
studied explicitly in section 8.

\section{Four-fermion models}

\setcounter{equation}{0}

In this section we extend our analysis to the four fermion models. Start
from the Euclidean four fermion lagrangian 
\begin{equation}
\mathcal{L}=\sum_{i=1}^{N}\overline{\psi }_{i}\left( \widehat{\partial }%
\!\!\!\slash+\frac{{\overline{\partial }\!\!\!\slash\,}^{n}}{\Lambda
_{L}^{n-1}}\right) \psi _{i}-\frac{\lambda ^{2}}{2\Lambda _{L}^{\overline{d}%
(1-1/n)}M^{\hbox{\dj }-2}}\left( \sum_{i=1}^{N}\overline{\psi }_{i}\psi
_{i}\right) ^{2},  \label{elle}
\end{equation}
for \hbox{\dj }$>2$. This model is not renormalizable by weighted power
counting, but, under certain conditions, it becomes renormalizable in the
large $N$ expansion. Introduce an auxiliary field $\sigma $ of weight 1 and
rewrite the lagrangian as 
\begin{equation}
\mathcal{L}=\sum_{i=1}^{N}\overline{\psi }_{i}\left( \widehat{\partial }%
\!\!\!\slash+\frac{{\overline{\partial }\!\!\!\slash\,}^{n}}{\Lambda
_{L}^{n-1}}+\lambda \sigma \right) \psi _{i}+\frac{\sigma ^{2}}{2}\Lambda
_{L}^{\overline{d}(1-1/n)}M^{\hbox{\dj }-2}\hbox{.}  \label{ey}
\end{equation}
In the large $N$ expansion the resummation of the bubble diagrams of Fig. 2
modifies the naive $\sigma $-propagator 
\[
\frac{1}{\Lambda _{L}^{\overline{d}(1-1/n)}M^{\hbox{\dj }-2}} 
\]
into 
\[
\frac{1}{\Lambda _{L}^{\overline{d}(1-1/n)}M^{\hbox{\dj }-2}+Q_{f}(\widehat{k%
},\overline{k},\Lambda _{L})}, 
\]
where $-Q_{f}(\widehat{k},\overline{k},\Lambda _{L})$ is the value of the
bubble diagram (see (\ref{fprop}) for an explicit expression in a concrete
case). Because of the mass $M$, super-renormalizable terms $\Delta _{sr}%
\mathcal{L}$ are generated by renormalization, proportional to integer
powers of $M^{\hbox{\dj }-2}$. For the moment we assume that no new strictly
renormalizable vertex is turned on and later determine the conditions under
which this effectively happens. Under these assumptions the complete
renormalizable lagrangian reads 
\begin{equation}
\mathcal{L}=\sum_{i=1}^{N}\overline{\psi }_{i}\left( \widehat{\partial }%
\!\!\!\slash+\frac{{\overline{\partial }\!\!\!\slash\,}^{n}}{\Lambda
_{L}^{n-1}}+\lambda \sigma \right) \psi _{i}+\frac{\sigma ^{2}}{2}\Lambda
_{L}^{\overline{d}(1-1/n)}M^{\hbox{\dj }-2}+\Delta _{sr}\mathcal{L}\hbox{.}
\label{ey2}
\end{equation}
At $M=0$ $Q_{f}$ is a homogeneous function of $\widehat{k}$ and $\overline{k}
$ and has the correct weight, equal to \hbox{\dj }$-2$, to ensure the
renormalizability of (\ref{ey2}) by weighted power counting in the large $N$
expansion (see (\ref{beha}) for an example), if we assume that the $\sigma $
propagator is regular (the conditions for its regularity are derived below).
Observe that the $\lambda $ beta function of (\ref{ey2}) vanishes
identically, since $\lambda $ is a redundant parameter that can be
reabsorbed in $\sigma $ and $M$. The renormalization constant of the vertex $%
\sigma \overline{\psi }\psi $ can be interpreted as the $\sigma $ wave
function renormalization constant.

Since $\Delta _{sr}\mathcal{L}$ vanishes at $M=0$, we see that in the
ultraviolet limit $M\rightarrow 0$ the four fermion theory (\ref{ey2}) flows
to the weighted scale invariant fixed point 
\begin{equation}
\mathcal{L}_{C}=\sum_{i=1}^{N}\overline{\psi }_{i}\left( \widehat{\partial }%
\!\!\!\slash+\frac{{\overline{\partial }\!\!\!\slash\,}^{n}}{\Lambda
_{L}^{n-1}}+\lambda \sigma \right) \psi _{i},  \label{LC}
\end{equation}
whose Minkowskian lagrangian reads 
\begin{equation}
\mathcal{L}_{\mathrm{M}}=\sum_{i=1}^{N}\overline{\psi }_{i}\left( i\widehat{%
\partial }\!\!\!\slash_{\mathrm{M}}+\frac{(i{\overline{\partial }\!\!\!\slash%
)}^{n}}{\Lambda _{L}^{n-1}}-\lambda \sigma _{\mathrm{M}}\right) \psi _{i},
\label{LCM}
\end{equation}
with $\sigma _{\mathrm{M}}=\sigma $. Observe that (\ref{ey2}) is an
interesting example of asymptotically safe theory \cite{asysafety}, its
interacting UV fixed point being indeed (\ref{LC})-(\ref{LCM}).

Let us study the properties of the fixed point. First observe that (\ref{LCM}%
) is invariant under parity: 
\[
P:\qquad \psi \rightarrow \gamma _{0}\psi ,\qquad \sigma \rightarrow \sigma
,\qquad x^{0}\rightarrow x^{0},\qquad x^{\mu }\rightarrow -x^{\mu }%
\hbox{
for }\mu \neq 0, 
\]
but when $n$ is odd the theory is invariant also under reflection $P_{(\mu
)} $ with respect to every space axis $\mu \neq 0$, precisely

\begin{equation}
P_{(\mu )}:\qquad \psi \rightarrow \gamma _{\mu }\psi ,\qquad \sigma
\rightarrow -\sigma ,\qquad x^{\mu }\rightarrow -x^{\mu },\qquad x^{\nu
}\rightarrow x^{\nu }\hbox{ for }\nu \neq \mu .  \label{reflection}
\end{equation}

\paragraph{Unitarity}

The positivity of the $\sigma $ spectral function is always guaranteed, for
an argument analogous to the one of the previous section. The fermion bubble
of Fig. 2 has weight equal to \hbox{\dj }$-2$, so $\rho (s,0)\sim 1/s^{a}$
with $a=($\hbox{\dj }$-2)/2$. The unitarity bound (\ref{a<1}) gives %
\hbox{\dj }$<4$, therefore we are going to study the models with 
\begin{equation}
2<\hbox{\dj }<4.  \label{bound}
\end{equation}

\paragraph{Renormalizability}

We now study the counterterms and impose that (\ref{LC}) be renormalizable
as it stands, in particular that no new strictly renormalizable term be
turned on. The conditions that we find ensure also the renormalizability of (%
\ref{ey2}). First, the counterterm 
\begin{equation}
\sigma \overline{\partial }^{q}\sigma  \label{forbid}
\end{equation}
must be forbidden for every non-negative even integer $q$. Its weight $2+q/n$
must be equal to \hbox{\dj }, so 
\[
q=n(\hbox{\dj }-2)=\overline{d}(1-n)+n(d-2). 
\]
Since \hbox{\dj }$>2$ we must have either $i$) $n$, $d$ both odd, or $ii$) $%
n $ even and $\overline{d}$ odd.

We want to forbid also 
\begin{equation}
\lbrack \widehat{\partial }^{p}\overline{\partial }^{q}]\sigma ^{3}
\label{tres}
\end{equation}
for every non-negative integers $p$, $q$ and every even $n$. Indeed, when $n$
is odd the counterterm is already forbidden by the invariance under (\ref
{reflection}). In the other cases we have 
\[
p+\frac{q}{n}=\hbox{\dj }-3. 
\]
The cases with \hbox{\dj }$<3$ are fine, because either $p$ or $q$ must be
negative. The case \hbox{\dj }$=3$ ($n$ even) is forbidden. Finally, when %
\hbox{\dj }$>3$, $n$ even implies $\overline{d}$ odd again.

It is easy to show that other terms such as 
\begin{equation}
\lbrack \widehat{\partial }^{p}\overline{\partial }^{q}]\sigma ^{m}
\label{soddo}
\end{equation}
are automatically forbidden, for every even non-negative integers $p$ and $q$
and for every $m\geq 4$.

Next, observe that there is a unique parity-invariant vertex with both $%
\sigma $ and fermion legs, that is $\sigma \overline{\psi }\psi $. It
remains to consider only the counterterms containing four or more fermions
and no $\sigma $. The four-fermion terms are symbolically written as 
\begin{equation}
\lbrack \widehat{\partial }^{p}\overline{\partial }^{q}][\overline{\psi }%
^{2}\psi ^{2}].  \label{ct}
\end{equation}
Every non-negative integers $p$ and $q$ must be excluded. Equating the
weight of (\ref{ct}) to \hbox{\dj } we have 
\[
np+q=n(2-\hbox{\dj }), 
\]
so these terms are automatically excluded for \hbox{\dj }$>2$. The exclusion
of counterterms of the form $[\widehat{\partial }^{p}\overline{\partial }%
^{q}][\overline{\psi }^{m}\psi ^{m}]$ with $m>2$ is also guaranteed.

Summarizing, we have non-renormalizable models (that become renormalizable
in the $1/N$ expansion) when 
\begin{equation}
2<\hbox{\dj }<4,\qquad \hbox{and either \textit{i}) or \textit{ii}) hold,}
\label{ff}
\end{equation}
where, again, \textit{i}) means that $n$, $d$ are both odd, and \textit{ii})
means that $n$ even and $\overline{d}$ is odd. However, the case \textit{\dj 
}$=3$ with $n$ even is excluded.

Let us list the four dimensional solutions to the conditions found so far.
We can have $\widehat{d}=1,2$ or $3$. When $\widehat{d}=1$, the unique
solution has $n=2$, \hbox{\dj }$=5/2$. When $\widehat{d}=2$ no solution is
admitted. Finally, when $\widehat{d}=3$ every even $n$ is a solution. The
simplest model of this class has $n=2$, \hbox{\dj }$=7/2$. It is studied
explicitly in section 9, where we show that it is stable and regular, and
calculate its subleading corrections.

Since the fermion bubble is always superficially divergent, in general
stability and regularity have to be studied case by case. Nevertheless,
there exists a noticeable class of odd-dimensional models that can be proved
to be stable and regular with a simple argument.

\paragraph{Stability and regularity for odd $n$}

Write $n=2m+1$ and set $\Lambda _{L}=1$ for simplicity. After a few
straightforward steps the Euclidean fermion bubble can be written as 
\[
2^{[d/2]-1}\lambda ^{2}N\int \frac{\widehat{p}^{2}+(\widehat{p}+\widehat{k}%
)^{2}-\widehat{k}^{2}\,+(\overline{p}^{2}+(\overline{p}+\overline{k})^{2}-%
\overline{k}^{2})ab}{\left( \widehat{p}^{2}+\overline{p}^{2}a^{2}\right)
\left( (\widehat{p}+\widehat{k})^{2}+(\overline{p}+\overline{k}%
)^{2}b^{2}\right) }, 
\]
where $a=(\overline{p}^{2})^{m}$, $b=\left( (\overline{p}+\overline{k}%
)^{2}\right) ^{m}$. We can replace $\widehat{p}^{2}$ with $-\overline{p}%
^{2}a^{2}$ and $(\widehat{p}+\widehat{k})^{2}$ with $-(\overline{p}+%
\overline{k})^{2}b^{2}$ in the numerator, since the difference is a tadpole
and vanishes identically using the dimensional-regularization technique. We
get 
\begin{equation}
-2^{[d/2]-1}\lambda ^{2}N\int \frac{\widehat{k}^{2}\,+\overline{k}^{2}ab+(%
\overline{p}^{2}a-(\overline{p}+\overline{k})^{2}b)(a-b)}{\left( \widehat{p}%
^{2}+\overline{p}^{2}a^{2}\right) \left( (\widehat{p}+\widehat{k})^{2}+(%
\overline{p}+\overline{k})^{2}b^{2}\right) }.  \label{cico}
\end{equation}
Now the numerator is positive definite and the integral is convergent for 
\begin{equation}
\hbox{\dj }<2+\frac{2}{n}\qquad \hbox{(}n\hbox{ odd).}  \label{nega}
\end{equation}
Therefore, when (\ref{nega}) holds the fermion bubble is negative-definite,
which implies that the $\sigma $ two-point function is positive definite and
the theory is guaranteed to be stable. For the same reason, setting $%
\overline{k}=0$ or $\widehat{k}=0$ in (\ref{cico}) gives the power-like
behaviors (\ref{uvbeha}) with $w=$\hbox{\dj }$-2$, proving regularity.

\bigskip

The simplest example of solutions to (\ref{nega}) and (\ref{ff}) is the
Lorentz invariant ($n=1$) four-fermion model in three spacetime dimensions 
\cite{parisi}. For $d>3$ the solutions must have $\widehat{d}=1$, since for $%
\widehat{d}>1$ (\ref{nega}) cannot be fulfilled. Then we find $d=n+2$. These
solutions generalize the four-fermion models of ref. \cite{parisi} to
arbitrary odd dimensions. Since $n$ is odd, the reflection symmetry $P_{(\mu
)}$ (\ref{reflection}) ensures that diagrams with an odd number of external $%
\sigma $-legs and no external $\psi $-leg vanish identically. Therefore, in
these models only the diagrams ($a$) and ($b$) of Fig. 3 contribute to the
renormalization group flow up to the next-to-leading order.

\section{Renormalization group}

\setcounter{equation}{0}

The fixed points of our Lorentz violating models are not conformal field
theories, but they are exactly weighted scale invariant. They depend on the
scale $\Lambda _{L}$ and have two correlation lengths (if the Lorentz group
is split into two subfactors, more otherwise). The symmetry under weighted
scale transformations is not sufficient to determine the two-point and
three-point functions up to a finite number of constants. In this section we
study the form of the two-point functions in the Lifshitz type models (\ref
{fix}). The treatment is general and applies to the fermion models of
section 6 with minor modifications.

We have wave-function renormalization constants for $\varphi $ and $\sigma $
and a renormalization constant for $\Lambda _{L}$. They are just functions
of $N$. The bare quantities are 
\begin{equation}
\varphi _{i\mathrm{B}}=Z_{\varphi }^{1/2}\varphi _{i},\qquad \Lambda _{%
\mathrm{B}L}=Z_{\Lambda }\Lambda _{L},\qquad \sigma _{\mathrm{B}}=\sigma
Z_{\sigma }^{1/2},\qquad \lambda _{\mathrm{B}}=\lambda .  \label{reno}
\end{equation}

The Callan-Symanzik equation reads 
\begin{equation}
\left( \mu \frac{\partial }{\partial \mu }+\eta _{L}\Lambda _{L}\frac{%
\partial }{\partial \Lambda _{L}}+k\gamma _{\varphi }+m\gamma _{\sigma
}\right) \left\langle \varphi (x_{1})\cdots \varphi (x_{k})\ \sigma
(y_{1})\cdots \sigma (y_{m})\right\rangle =0,  \label{CS}
\end{equation}
where 
\[
\gamma _{\varphi }=\frac{1}{2}\frac{\mathrm{d}\ln Z_{\varphi }}{\mathrm{d}%
\ln \mu },\qquad \gamma _{\sigma }=\frac{1}{2}\frac{\mathrm{d}\ln Z_{\sigma }%
}{\mathrm{d}\ln \mu },\qquad \eta _{L}=-\frac{\mathrm{d}\ln Z_{\Lambda }}{%
\mathrm{d}\ln \mu }. 
\]

Consider the two-point function $G(|\widehat{x}|,|\overline{x}|;N,\Lambda
_{L},\mu )\equiv \left\langle \varphi (x)\ \varphi (0)\right\rangle $.
Because of the residual Lorentz invariance $O(1,\widehat{d}-1)\otimes O(%
\overline{d})$ the correlation function depends only on $|\widehat{x}|,|%
\overline{x}|$. The RG equations tell us that 
\begin{equation}
G(|\widehat{x}|,|\overline{x}|;N,\xi ^{\eta _{L}}\Lambda _{L},\xi \mu )=\xi
^{-2\gamma _{\varphi }}G(|\widehat{x}|,|\overline{x}|;N,\Lambda _{L},\mu ).
\label{i1}
\end{equation}
On the other hand, the invariance with respect to (\ref{scale2}) gives 
\begin{equation}
G(\xi |\widehat{x}|,\xi ^{1/n}|\overline{x}|;N,\Lambda _{L},\xi ^{-1}\mu
)=\xi ^{2-\hbox{\dj }}G(|\widehat{x}|,|\overline{x}|;N,\Lambda _{L},\mu ).
\label{i2}
\end{equation}
Finally, dimensional analysis gives 
\begin{equation}
G(\xi |\widehat{x}|,\xi |\overline{x}|;N,\xi ^{-1}\Lambda _{L},\xi ^{-1}\mu
)=\xi ^{2-d}G(|\widehat{x}|,|\overline{x}|;N,\Lambda _{L},\mu ).  \label{i3}
\end{equation}
There is only one dimensionless combination of $|\widehat{x}|$, $|\overline{x%
}|$, $\Lambda _{L}$ and $\mu $ that is RG invariant and invariant under (\ref
{scale2}), namely 
\[
\frac{|\widehat{x}|}{|\overline{x}|^{n}\Lambda _{L}^{n-1}}(|\widehat{x}|\mu
)^{\eta _{L}(n-1)}, 
\]
therefore the solution contains an arbitrary function $G_{r}$ of it. The
other dependencies can be fixed straightforwardly and the result is 
\[
G(|\widehat{x}|,|\overline{x}|;\lambda ,\Lambda _{L},\mu )=\frac{1}{|%
\widehat{x}|^{\hbox{\dj }-2}(\mu |\widehat{x}|)^{2\gamma _{\varphi }}}\left( 
\frac{|\overline{x}|^{n}}{|\widehat{x}|}\right) ^{(\hbox{\textit{\dj }}%
-d)/(n-1)}G_{r}\left( N,\frac{|\widehat{x}|}{|\overline{x}|^{n}\Lambda
_{L}^{n-1}}(|\widehat{x}|\mu )^{\eta _{L}(n-1)}\right) . 
\]
Even simpler is the form of the two-point function in momentum space, which
is 
\[
\widetilde{G}(|\widehat{p}|,|\overline{p}|;\lambda ,\Lambda _{L},\mu )=\frac{%
1}{|\widehat{p}|^{2(1-\gamma _{\varphi })}\mu ^{2\gamma _{\varphi }}}%
\widetilde{G}_{r}^{\prime }\left( N,\frac{|\widehat{p}|\Lambda _{L}^{n-1}}{|%
\overline{p}|^{n}}\left( \frac{|\widehat{p}|}{\mu }\right) ^{\eta
_{L}(n-1)}\right) . 
\]
From the zeroth order propagator we have 
\[
\widetilde{G}_{r}^{\prime }\left( \infty ,u\right) =\frac{u^{2}}{1+u^{2}}. 
\]

\section{Fixed points of scalar theories}

\setcounter{equation}{0}

In this section we give results about the four dimensional Lifshitz type
fixed point (\ref{fix}) with $\widehat{d}=3$, $\overline{d}=1$, $n=2$, \dj 
\textit{=}$7/2$ up to the subleading order in the $1/N$ expansion. The model
is perturbatively unitary, causal, stable and regular.

The scalar bubble (Fig. 2) is evaluated in appendix A. The $\sigma $%
-propagator has weight $1/2$ and it is equal to minus the reciprocal of (\ref
{bub}), namely 
\begin{equation}
\frac{16\pi |\widehat{k}|\sqrt{2}}{\lambda ^{2}N\sqrt{F-\overline{k}^{2}}}
\label{sigmapropa}
\end{equation}
where $F=\sqrt{\overline{k}^{4}+4\widehat{k}^{2}\Lambda _{L}^{2}}$. The
positivity of (\ref{sigmapropa}), which guarantees stability, is a direct
consequence of the superficial renormalizability of the scalar bubble. The
asymptotic behaviors in the limits $|\widehat{k}|\Lambda _{L}\gg \overline{k}%
^{2}$ and $|\widehat{k}|\Lambda _{L}\ll \overline{k}^{2}$, namely 
\begin{equation}
\frac{16\pi }{\lambda ^{2}N}\sqrt{\frac{|\widehat{k}|}{\Lambda _{L}}},\qquad 
\frac{16\pi |\overline{k}|}{\lambda ^{2}N\Lambda _{L}},  \label{beha}
\end{equation}
agree with the regularity conditions (\ref{uvbeha}) in the ultraviolet
limits $\widehat{k}\rightarrow \infty $ and $\overline{k}\rightarrow \infty $%
, respectively, and also prove smoothness in the infrared regions $\overline{%
k}\rightarrow 0$ and $\widehat{k}\rightarrow 0$, which guarantees the
absence of spurious IR\ divergences. Finally, it is evident that the
propagator (\ref{sigmapropa}) is manifestly regular everywhere else.

We now use the propagator (\ref{sigmapropa}) to compute the subleading
corrections.

\paragraph{Subleading corrections}

\begin{figure}[tbp]
\centerline{\includegraphics[width=4.5in,height=1in]{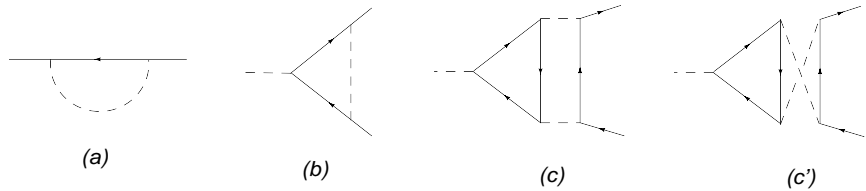}}
\caption{Subleading corrections}
\end{figure}
The $1/N$ corrections can be worked out computing the divergent parts of the
diagrams $(a)$, $(b)$ and $(c)$ drawn in Fig. 3. The dashed line is the $%
\sigma $-propagator (\ref{sigmapropa}). The continuous line denotes the
scalar field. The orientation of the continuous line is immaterial in the
scalar case, so the diagram $(c^{\prime })$ is the same as $(c)$ and should
not be counted. In the Lorentz-invariant three-dimensional models \cite
{arefeva} the diagram $(c)$ is convergent, but in the four-dimensional model
that we are considering now it is not. Nevertheless, the calculation of its
divergent part can be luckily carried over to the very end, using the
strategy illustrated in Appendix E. The results are 
\begin{equation}
(a)=\frac{1}{9\pi N\varepsilon \sqrt{3}}\left( -5\widehat{p}^{2}+\frac{1}{3}%
\frac{(\overline{p}^{2})^{2}}{\Lambda _{L}^{2}}\right) ,\qquad (b)=\frac{%
4i\lambda }{\pi N\varepsilon \sqrt{3}},\qquad (c)=\frac{14i\lambda }{3\pi
N\varepsilon \sqrt{3}},  \label{dds}
\end{equation}
whence 
\begin{equation}
\gamma _{\varphi }=\frac{5}{18\pi N\varepsilon \sqrt{3}},\qquad \eta _{L}=-%
\frac{8}{27\pi N\sqrt{3}},\qquad \gamma _{\sigma }=-\frac{83}{9\pi N\sqrt{3}}%
.  \label{f1}
\end{equation}
The results (a), $\gamma _{\varphi }$ and $\eta _{L}$ agree with the ones
found in ref. \cite{largeNlifshitz}, while (b), (c) and $\gamma _{\sigma }$
are new.

\section{Fixed points of fermion theories}

\setcounter{equation}{0}

With simple generalizations the techniques used in the previous section
apply also to the four-dimensional fermionic model (\ref{LC}) with $\widehat{%
d}=3$, $\overline{d}=1$, $n=2$, \hbox{\dj }$=7/2$. This model is
perturbatively causal, unitary, stable and regular. Since the $\sigma $
bubble is superficially divergent stability and regularity can be proved
only by explicit computation, done in appendix B.

The $\sigma $-propagator has weight $-3/2$ and can be read from (\ref{fprop}%
): 
\begin{equation}
\frac{15\pi \left( \overline{k}^{2}+F\right) ^{3/2}}{\lambda ^{2}N\sqrt{2}%
\Lambda _{L}\widehat{k}^{2}\left( 2\overline{k}^{2}+3F\right) }.  \label{pos}
\end{equation}
Its positivity proves stability. The asymptotic behavior for $|\widehat{k}%
|\Lambda _{L}\gg \overline{k}^{2}$, which is 
\begin{equation}
\frac{5\pi }{\lambda ^{2}N|\widehat{k}|\sqrt{\Lambda _{L}|\widehat{k}|}},
\label{beha2}
\end{equation}
agrees with the regularity condition (\ref{uvbeha}) for $\widehat{k}%
\rightarrow \infty $ and, at the same time, proves smoothness for $\overline{%
k}\rightarrow 0$. The asymptotic behavior for $|\widehat{k}|\Lambda _{L}\ll 
\overline{k}^{2}$, instead, 
\begin{equation}
\frac{6\pi |\overline{k}|}{\lambda ^{2}N\Lambda _{L}\widehat{k}^{2}}
\label{beha3}
\end{equation}
does not agree with (\ref{uvbeha}) and deserves more attention. When $%
\overline{k}\rightarrow \infty $ and $\widehat{k}$ is kept fixed (or grows
more slowly than $\overline{k}$) the behavior (\ref{beha3}) could generate a
spurious UV ``sub''divergence in the integral over $\overline{k}$. Now we
prove that it is not so. Consider a diagram $G$ with integrated momenta $k$, 
$L$ loops, $V$ vertices, $I=I_{\sigma }+I_{\psi }$ internal legs and $%
E=E_{\sigma }+E_{\psi }$ external legs. Using $L=I-V+1$, $E_{\sigma
}+2I_{\sigma }=V$ and $E_{\psi }+2I_{\psi }=2V$, we get $L=1+I_{\sigma
}-E_{\psi }/2$. Since, $L\geq 1$ we obtain the bound $E_{\psi }\leq
2I_{\sigma }$. Each fermion propagator behaves like $1/\overline{k}^{2}$, so
the degree of divergence $\overline{\omega }(G)$ of the subintegral over $%
\overline{k}$ is 
\[
\overline{\omega }(G)=L+I_{\sigma }-2I_{\psi }=1-2I_{\sigma }+\frac{E_{\psi }%
}{2}-2E_{\sigma }\leq 1-I_{\sigma }-2E_{\sigma }\hbox{.}
\]
Spurious UV\ divergences ($\overline{\omega }\geq 0$) can occur only for $%
E_{\sigma }=0$, $I_{\sigma }=1$, which implies $E_{\psi }\leq 2$. The unique
diagram with a potential problem is the one-loop fermion self energy (a).
However, because of (\ref{beha3}) in that case the potentially dangerous
behavior reads 
\[
\int \frac{\mathrm{d}^{3}\widehat{k}}{(2\pi )^{3}}\frac{1}{\widehat{k}^{2}}%
\int_{|\overline{k}|\sim \infty }\frac{\mathrm{d}\overline{k}}{2\pi }\frac{1%
}{|\overline{k}|}.
\]
While the $\overline{k}$-integral is logarithmic divergent, it is multiplied
by a $\widehat{k}$-integral that vanishes identically in dimensional
regularization. Thus the behavior (\ref{beha3}) is not dangerous for $%
\overline{k}\rightarrow \infty $.

Finally, when $\widehat{k}\rightarrow 0$ the behavior (\ref{beha3})
guarantees that no spurious IR divergences affects the $\widehat{k}$%
-integral at non-exceptional external hatted momenta. In every other region
the $\sigma $ propagator (\ref{pos}) is manifestly regular.

\paragraph{Subleading corrections}

The contributing diagrams are $(a)$, $(b)$, $(c)$ and $(c^{\prime })$ of
Fig. 2. We find 
\begin{eqnarray}
(a) &=&-\frac{1}{3844\pi N\varepsilon \sqrt{3}}\left( i\widehat{p}\!\!\!%
\slash (5085-206\sqrt{10})+\frac{3}{31}(47625-12146\sqrt{10})\frac{\overline{%
p}^{2}}{\Lambda _{L}}\right) ,  \nonumber \\
(b) &=&\frac{\lambda \sqrt{3}(495+82\sqrt{10})}{3844\pi N\varepsilon }%
,\qquad \qquad (c)+(c^{\prime })=\frac{\lambda \sqrt{3}}{961\pi N\varepsilon 
}(900-113\sqrt{10}),  \label{ddf}
\end{eqnarray}
whence 
\begin{equation}
\gamma _{\psi }=\frac{5085-206\sqrt{10}}{7688\pi N\sqrt{3}},\qquad \eta
_{L}=-\frac{53(2835-404\sqrt{10})}{59582\pi N\sqrt{3}},\qquad \gamma
_{\sigma }=-\frac{8685-658\sqrt{10}}{1922\pi N\sqrt{3}}.  \label{f2}
\end{equation}

\section{RG interpolation between pairs of fixed points}

\setcounter{equation}{0}

Following ref.s \cite{erg} and \cite{erg2} in this section we construct RG
flows interpolating between pairs of fixed points of the types studied in
sections 8 and 9.

The interpolating theories have lagrangians 
\begin{eqnarray*}
\mathcal{L}_{\varphi ,\phi } &=&\frac{1}{2}\sum_{i=1}^{N}\left[ (\widehat{%
\partial }\varphi _{i})^{2}+\frac{1}{\Lambda _{L}^{2}}(\overline{\partial }%
^{2}\varphi _{i})^{2}+i\sigma \varphi _{i}^{2}\right] +\frac{1}{2}%
\sum_{j=1}^{M}\left[ (\widehat{\partial }\phi _{j})^{2}+\frac{f^{2}}{\Lambda
_{L}^{2}}(\overline{\partial }^{2}\phi _{j})^{2}+ig\sigma \phi
_{j}^{2}\right] , \\
\mathcal{L}_{\psi ,\chi } &=&\sum_{i=1}^{N}\overline{\psi }_{i}\left( 
\widehat{\partial }\!\!\!\slash+\frac{\overline{\partial }^{2}}{\Lambda _{L}}%
+\sigma \right) \psi _{i}+\sum_{j=1}^{M}\overline{\chi }_{j}\left( \widehat{%
\partial }\!\!\!\slash+f\frac{\overline{\partial }^{2}{\,}}{\Lambda _{L}}%
+g\sigma \right) \chi _{j}.
\end{eqnarray*}
It is straightforward to prove the renormalizability of such models. The
only caveat, with respect to the analysis of counterterms performed section
5 (for scalar fields) is that we must exclude also new counterterms
proportional to $\sum_{i=1}^{N}\varphi _{i}^{2}$, because they are no longer
proportional to the $\sigma $ field equations. Since \dj \textit{=}$7/2$ the
weight of $\varphi $, $\phi $ is $3/4$, so counterterms with four or more
scalars are forbidden by locality.

In practice, the interpolating theories are made by pairs of models (\ref
{fix}) or (\ref{LC}) sharing the same field $\sigma $. Here the parameter $%
\lambda $ is suppressed (reabsorbed inside $\sigma $) and the running
couplings\ are $f$ and $g$. The fixed points are the zeros of the $f$ and $g 
$ beta functions. There is an evident duality 
\begin{equation}
g\leftrightarrow \frac{1}{g},\qquad f\leftrightarrow \frac{1}{f},\qquad
\Lambda _{L}\leftrightarrow \frac{\Lambda _{L}}{f},\qquad N\leftrightarrow M.
\label{dual}
\end{equation}

The phase diagrams contain some remarkable fixed points. When $g\rightarrow
0 $ the models tend to the fixed points (\ref{fix}) or (\ref{LC}) plus some
free fields. When $f,g\rightarrow 1$ the models tend to the selfdual fixed
points (\ref{fix}) or (\ref{LC}) with $N\rightarrow N+M$. Because of the
duality (\ref{dual}), when $g\rightarrow \infty $ the models tend to the
fixed points (\ref{fix}) or (\ref{LC}) with $N\rightarrow M$, plus free
fields. The phase diagrams might contain also some new fixed points.

The bubble diagrams of the interpolating models can be easily calculated
using the results of the fixed points. We have the sum of two terms: the
first contribution is due to circulating $\varphi $,$\psi $ fields and
coincides with (\ref{bub}) or (\ref{fprop}) (at $\lambda =1$); the second
contribution is due to circulating $\phi $,$\chi $ fields and is equal to $%
rg^{2}$ times (\ref{bub}) or (\ref{fprop}), but with $\Lambda _{L}$ replaced
by $\Lambda _{L}/f$, where $r=M/N$. In total we have 
\[
-Q(\widehat{k},\overline{k},\Lambda _{L})-rg^{2}Q(\widehat{k},\overline{k}%
,\Lambda _{L}/f).
\]
This formula proves stability and regularity in both interpolating models.

Only finitely many graphs contribute at each order of the large $N$
expansion, so the subleading corrections can be calculated exactly in $f$, $%
g $ at every order in $1/N$.

\section{Conclusions}

\setcounter{equation}{0}

In this paper we have studied several properties of Lorentz violating
quantum field theories of scalars and fermions and constructed fixed points
of renormalization-group flows using a large $N$ expansion. Such fixed
points have an exact weighted scale invariance and are the best
generalizations of conformal field theories when the Lorentz symmetry is
violated.

Unitarity, causality and stability can be generalized straightforwardly to
Lorentz violating theories, because, strictly speaking, none of these
notions demands Lorentz invariance. We have classified the models that are
guaranteed to be unitary, causal and stable. In other models stability needs
to be verified explicitly case by case. Solutions exist also in four and
higher dimensions, while Lorentz invariant models of this type are known to
exist only in lower dimensions. This makes our new models potentially
interesting for applications to high-energy physics.

In some fixed points the calculations can be analytically carried over up to
the subleading corrections. Using a simple trick it is also easy to
construct running quantum field theories that interpolate between pairs of
fixed points.

The models constructed here and in ref. \cite{rennolor} enlarge considerably
the realm of renormalizable theories. They have a variety of potential
physical applications and provide a large laboratory to test ideas about
quantum field theory and renormalization.

\vskip 12truept \noindent {\large \textbf{Appendix A: Scalar bubble}}

\vskip 2truept

\renewcommand{\theequation}{A.\arabic{equation}} \setcounter{equation}{0}

Since some calculations in Lorentz violating theories have unusual aspects
we collect details and results in these appendices.

We study the $\sigma $ two-point function for the model (\ref{fix}) with $%
\widehat{d}=3$, $\overline{d}=1$, $n=2$, \textit{\dj =}$7/2$ at the leading
order in $1/N$. We first evaluate the bubble diagram of fig. 1 in the
Euclidean framework. Later we compute the imaginary part in the Minkowskian
framework using the cutting method. The graph reads

\begin{equation}
-\frac{\lambda ^{2}N}{2}\int \frac{\mathrm{d}^{3}\widehat{p}}{(2\pi )^{3}}%
\int_{-\infty }^{+\infty }\frac{\mathrm{d}\overline{p}}{2\pi }\frac{1}{%
\left( \widehat{p}^{2}+\frac{\overline{p}^{4}}{\Lambda _{L}^{2}}\right)
\left( \left( \widehat{p}-\widehat{k}\right) ^{2}+\frac{(\overline{p}-%
\overline{k})^{4}}{\Lambda _{L}^{2}}\right) }.  \label{sbubble}
\end{equation}
We integrate over $\widehat{p}$ using Feynman parameters and find 
\[
-\frac{\lambda ^{2}N}{8\pi |\widehat{k}|}\int_{-\infty }^{+\infty }\frac{%
\mathrm{d}\overline{p}}{2\pi }\arctan \frac{\Lambda _{L}|\widehat{k}|}{%
\overline{p}^{2}+(\overline{p}-\overline{k})^{2}}. 
\]
The integrand can be conveniently expanded in powers of $\widehat{k}^{2}$.
Using 
\begin{equation}
\int_{-\infty }^{+\infty }\frac{\mathrm{d}\overline{p}}{2\pi }\frac{1}{%
\left( \overline{p}^{2}+(\overline{p}-\overline{k})^{2}\right) ^{m}}=\frac{%
2^{m-2}\Gamma \left( m-\frac{1}{2}\right) }{\pi ^{1/2}|\overline{k}%
|^{2m-1}\Gamma (m)},\qquad m>\frac{1}{2},  \label{integra}
\end{equation}
and resumming the series, we arrive at 
\begin{equation}
-\frac{\lambda ^{2}N}{16\pi |\widehat{k}|\sqrt{2}}\sqrt{\sqrt{\overline{k}%
^{4}+4\widehat{k}^{2}\Lambda _{L}^{2}}-\overline{k}^{2}}\equiv -Q(\widehat{k}%
,\overline{k},\Lambda _{L}),  \label{bub}
\end{equation}
which agrees with the result of \cite{largeNlifshitz}. The contribution to
the generating functional $\Gamma $ of one-particle irreducible diagrams is
positive definite: 
\begin{equation}
\int \frac{\mathrm{d}^{4}k}{(2\pi )^{4}}\frac{1}{2}\widetilde{\sigma }(-k)Q(%
\widehat{k},\overline{k},\Lambda _{L})\widetilde{\sigma }(k)\geq 0,
\label{aga}
\end{equation}
in agreement with stability.

We now rotate the correlation function to the Minkowskian framework and
study the imaginary part of the scalar bubble. The lagrangian (\ref{fix}) is
turned into (\ref{minko}). We find a cut on the real axis for $\widehat{k}_{%
\hbox{M}}^{2}\geq \overline{k}^{4}/(4\Lambda _{L}^{2})$. The imaginary part
of the $\sigma _{\hbox{M}}$ bubble multiplied by $-2i$ results

\begin{equation}
\frac{\lambda ^{2}N}{16\pi \sqrt{\widehat{k}_{\hbox{M}}^{2}}}\theta \left( 
\widehat{k}_{\hbox{M}}^{2}-\frac{\overline{k}^{4}}{4\Lambda _{L}^{2}}\right) 
\sqrt{2\Lambda _{L}\sqrt{\widehat{k}_{\hbox{M}}^{2}}-\overline{k}^{2}}
\label{ima}
\end{equation}
and can be checked also directly computing the cutting diagram of the $%
\sigma $ self energy.

\vskip 12truept \noindent {\large \textbf{Appendix B: Fermion bubble}}

\vskip 2truept

\renewcommand{\theequation}{B.\arabic{equation}} \setcounter{equation}{0}

In this appendix we study the $\sigma $-two-point function of the model (\ref
{LC}) with $\widehat{d}=3$, $\overline{d}=1$, $n=2$, \hbox{\dj }$=7/2$ at
the leading order in $1/N$. The strategy of the calculation is the same as
in the previous section. The main difference with respect to the scalar case
is that now the integral is formally divergent. The divergence is however
power-like and so vanishes using the dimensional-regularization technique.
First we integrate over the momentum $\widehat{p}$ using standard
techniques. The result is 
\[
-\frac{\lambda ^{2}N}{2\pi }\int_{-\infty }^{+\infty }\frac{\mathrm{d}^{%
\overline{D}}\overline{p}}{2\pi }\left[ a+b+\frac{1}{|\widehat{k}|}\left( 
\widehat{k}^{2}+(a+b)^{2}\right) \arctan \frac{|\widehat{k}|}{a+b}\right]
,\qquad a=\frac{\overline{p}^{2}}{\Lambda _{L}},\qquad b=\frac{(\overline{p}-%
\overline{k})^{2}}{\Lambda _{L}}. 
\]
Next, it is convenient to expand the integrand in powers of $|\widehat{k}|$
and integrate the series term-by-term. The zeroth order term of the
expansion, equal to $2(a+b)$, is killed by the dimensional integral over $%
\overline{p}$. The integral of every other term is convergent in $\overline{D%
}=1$ and can be calculated again using (\ref{integra}). Resumming the series
we obtain 
\begin{equation}
-\frac{\lambda ^{2}N\sqrt{2}}{15\pi }\Lambda _{L}\widehat{k}^{2}\frac{2%
\overline{k}^{2}+3F}{\left( \overline{k}^{2}+F\right) ^{3/2}}\equiv -Q_{f}(%
\widehat{k},\overline{k},\Lambda _{L}).  \label{fprop}
\end{equation}

Again, the contribution to the generating functional $\Gamma $ of
one-particle irreducible diagrams is positive definite, 
\[
\int \frac{\mathrm{d}^{4}k}{(2\pi )^{4}}\frac{1}{2}\widetilde{\sigma }%
(-k)Q_{f}(\widehat{k},\overline{k},\Lambda _{L})\widetilde{\sigma }(k), 
\]
which proves stability.

The imaginary part of the Minkowskian $\sigma $ bubble multiplied by $-2i$
is 
\[
\frac{\lambda ^{2}N}{30\pi \Lambda _{L}^{2}\sqrt{\widehat{k}_{\mathrm{M}}^{2}%
}}\theta \left( \widehat{k}_{\mathrm{M}}^{2}-\frac{\overline{k}^{4}}{%
4\Lambda _{L}^{2}}\right) \left( 2\Lambda _{L}\sqrt{\widehat{k}_{\mathrm{M}%
}^{2}}-\overline{k}^{2}\right) ^{3/2}\left( 3\Lambda _{L}\sqrt{\widehat{k}_{%
\mathrm{M}}^{2}}+\overline{k}^{2}\right) . 
\]
Being positive definite, this result is a check that the theory is
perturbatively unitary.

\vskip 12truept \noindent {\large \textbf{Appendix C: Scalar triangle}}

\vskip 2truept

\renewcommand{\theequation}{C.\arabic{equation}} \setcounter{equation}{0}

In this section we compute the scalar triangle with one vanishing external
momentum. The triangle is necessary to compute the divergent part of the
diagram $(c)$. Denote the external momentum with $k$ and the loop momentum
with $p$. To avoid IR\ problems, we add a mass changing the scalar
propagator as follows: 
\begin{equation}
\frac{1}{\widehat{p}^{2}+(\overline{p}^{2}+m^{2})^{2}/\Lambda _{L}^{2}}.
\label{appro}
\end{equation}
The diagram is ultraviolet convergent. First we integrate over $\widehat{p}$
using Feynman parameters, then over $\overline{p}$. The $\widehat{p}$%
-integral gives 
\[
\frac{i\lambda ^{3}N}{8\pi a\left[ \widehat{k}^{2}+(a+b)^{2}\right] },\qquad %
\hbox{with\qquad }a=\frac{\overline{p}^{2}+m^{2}}{\Lambda _{L}},\qquad b=%
\frac{(\overline{p}-\overline{k})^{2}+m^{2}}{\Lambda _{L}}. 
\]
Next, the integral over $\overline{p}$ is easily done using the residue
theorem. The result can be expanded in powers of $m$ as 
\begin{equation}
\frac{i\lambda ^{3}N\Lambda _{L}^{3}}{16\pi |m|\left[ \widehat{k}^{2}\Lambda
_{L}^{2}+\left( \overline{k}^{2}+m^{2}\right) ^{2}\right] }+\frac{i\lambda
^{3}N\Lambda _{L}^{3}\left[ \left( \overline{k}^{4}-\Lambda _{L}^{2}\widehat{%
k}^{2}\right) \sqrt{F+\overline{k}^{2}}-2|\widehat{k}|\Lambda _{L}\overline{k%
}^{2}\sqrt{F-\overline{k}^{2}}\right] }{8\pi \sqrt{2}\sqrt{4\widehat{k}%
^{2}\Lambda _{L}^{2}+\left( \overline{k}^{2}+m^{2}\right) ^{2}}\left[ 
\widehat{k}^{2}\Lambda _{L}^{2}+\left( \overline{k}^{2}+m^{2}\right)
^{2}\right] ^{2}},  \label{st}
\end{equation}
up to $\mathcal{O}(m)$, where $F$ is defined in section 8. The masses in the
denominators have been left, again, to avoid IR problems in the calculation
of the two-loop diagram $(c)$, which is performed in Appendix E.

\vskip 12truept \noindent {\large \textbf{Appendix D: Fermion triangle}}

\vskip 2truept

\renewcommand{\theequation}{D.\arabic{equation}} \setcounter{equation}{0}

Now we calculate the fermion triangle with one vanishing external momentum.
The integral has no IR problem, so we do not need to introduce a mass. On
the other hand, the diagram is formally ultraviolet divergent. The UV
divergence is linear, so it vanishes in dimensional regularization.
Equivalently, we can work in the physical dimension and subtract an
appropriate local term. Again, the integration over the loop momentum $p$ is
first done over $\widehat{p}$ using Feynman parameters, then over $\overline{%
p}$. Call $k$ the external momentum. The $\widehat{p}$-integral gives 
\[
\frac{\lambda ^{3}N}{2\pi ^{2}|\widehat{k}|\Lambda _{L}}\int_{-\infty
}^{+\infty }\mathrm{d}\overline{p}\ \left[ \left( \overline{p}^{2}+(%
\overline{p}-\overline{k})^{2}\right) \arctan \frac{|\widehat{k}|\Lambda _{L}%
}{\overline{p}^{2}+(\overline{p}-\overline{k})^{2}}\right] . 
\]
The integral over $\overline{p}$ is easily done expanding the arctangent in
powers of its argument, eliminating the first contribution to the sum (which
subtracts the UV divergence) and using (\ref{integra}). Finally, resumming
the series back, we find the result 
\[
-\frac{\lambda ^{3}N\sqrt{2}\widehat{k}^{2}\Lambda _{L}^{2}}{3\pi \left( 
\overline{k}^{2}+F\right) ^{3/2}}. 
\]

\vskip 12truept \noindent {\large \textbf{Appendix E: Calculation of the
diagrams (a), (b) and (c)}}

\vskip 2truept

\renewcommand{\theequation}{E.\arabic{equation}} \setcounter{equation}{0}

Here we describe the strategy to calculate the divergent parts of the
diagrams (a), (b) and (c) of Fig. 3. We begin with the scalar self energy
(a). By locality, $O(1,\widehat{d}-1)\otimes O(\overline{d})$ invariance and
weighted power counting, its divergent part is parametrized as 
\[
a\widehat{k}^{2}+b\frac{(\overline{k}^{2})^{2}}{\Lambda _{L}^{2}}. 
\]
The constants $a$ and $b$ are calculated appropriately differentiating with
respect to $\widehat{k}$ and $\overline{k}$ and later setting $k=0$. To
avoid spurious IR divergences at $k=0$ it is useful to introduce a small
mass, e.g. using modified propagators of the form (\ref{appro}). The
coefficients $a$ and $b$ can be expressed as integrals 
\[
\int \frac{\mathrm{d}^{3}\widehat{q}}{(2\pi )^{3}}\int_{-\infty }^{+\infty }%
\frac{\mathrm{d}\overline{q}}{2\pi }\mathcal{F}(\widehat{q}^{2},\overline{q}%
^{2}+m^{2}), 
\]
where $\mathcal{F}(\widehat{q}^{2},\overline{q}^{2}+m^{2})$ is a certain
homogeneous function of degree $-7/2$ (giving a weight $1/2$ to $m$).
Calling $x=\widehat{q}^{2}\Lambda _{L}^{2}/(\overline{q}^{2}+m^{2})^{2}$ and
changing variables from \mbox{$\vert$}$\widehat{q}|$ to $x$ we get an
expression of the form 
\begin{equation}
\int^{\pm \infty }\frac{\mathrm{d}\overline{q}}{\sqrt{\overline{q}^{2}+m^{2}}%
}\int_{0}^{\infty }\mathrm{d}x\ \mathbf{f}(x),  \label{essa}
\end{equation}
where $\mathbf{f}(x)$ is a function whose integral over $x$ is convergent.
The factorized integral over $\overline{q}$ gives instead the logarithmic
divergence. Observe that in dimensional regularization the divergences are
poles in $\varepsilon =\varepsilon _{1}+\varepsilon _{2}/2$ \cite{rennolor},
where $\widehat{d}-\varepsilon _{1}$ and $\overline{d}-\varepsilon _{2}$ are
the complex continuations of the dimensions $\widehat{d}$ and $\overline{d}$%
. In our calculation $\varepsilon _{1}$ can be kept equal to zero to the
very end, while the integral over $\overline{q}$ needs to be continued in
order to extract its UV divergence. We have 
\[
\int \frac{\mathrm{d}^{1-\varepsilon _{2}}\overline{q}}{\sqrt{\overline{q}%
^{2}+m^{2}}}=\frac{2}{\varepsilon _{2}}+\hbox{finite}=\frac{1}{\varepsilon }+%
\hbox{finite}, 
\]
so in (\ref{essa}) the residue of the pole is the value of the $x$-integral
of $\mathbf{f}(x)$.

The same strategy is used to calculate the other diagrams, both in the
scalar and fermion models. Basically, a logarithmic divergent integral over $%
\overline{q}$ is factored out. It multiplies a convergent integral over $x$
that can be evaluated exactly. In the case of fermions no IR divergence
occurs at vanishing external momenta, so there is no need to introduce the
mass $m$.

Following these guidelights, the calculations proceed straightforwardly in
all cases but one: the diagram (c) for scalar fields. As usual, its
divergent part can be worked out setting its external momenta to zero and
introducing an auxiliary mass. The triangle diagram with one vanishing
external momentum, calculated in Appendix C, is sufficient for the
evaluation. We know that the scalar triangle is IR\ divergent, as shown by
formula (\ref{st}). The first term of (\ref{st}), however, vanishes when
inserted in the rest of (c) and integrated. Indeed, this operation gives a
result of the form $I/|m|$, where $I$ is the value of some integral of
weight $1/2$. By weighted power counting, $I$ can only diverge linearly, and
by locality its divergence can only have the form $m/\varepsilon $ (not $%
|m|/\varepsilon $ !). Therefore the divergent contribution due to the first
term of (\ref{st}) would depend on the sign of $m$, which is absurd, since $%
m $ is introduced via the modified propagator (\ref{appro}). So, it is
sufficient to keep the second term of (\ref{st}). Once that term is inserted
in the rest of (c) its manipulation is straightforward following the
strategy described for diagram (a).

\end{document}